\documentclass{ws-procs975x65}%
\usepackage{amsmath}
\usepackage{amsfonts}
\usepackage{amssymb}
\usepackage{graphicx}%
\def\beq{\begin{equation}}
\def\eeq{\end{equation}}

\begin{document}

\title{Gravity's Rainbow and Compact Stars}

\author{Remo Garattini}

\address{Universit\`{a} degli Studi di Bergamo, \\
Dipartimento di Ingegneria e scienze applicate,\\
Viale Marconi,5 24044 Dalmine (Bergamo) ITALY\\
I.N.F.N. - sezione di Milano, Milan, Italy\\
E-mail:remo.garattini@unibg.it}

\author{Gianluca Mandanici}

\address{Universit\`{a} degli Studi di Bergamo, \\
Dipartimento di Ingegneria e scienze applicate,\\
Viale Marconi,5 24044 Dalmine (Bergamo) ITALY\\
E-mail:gianluca.mandanici@unibg.it}

\begin{abstract}
A number of recent studies has focused on the implications of new physics at
the Planck scale on the equilibrium of compact astrophysical objects such as
white dwarf and neutron stars. Here we analyze the modification of the
equilibrium configurations induced by the so-called Gravity's Rainbow that
account for Planck scale deformation of the space-time.

\end{abstract}

\bodymatter

\section{Introduction}

Compact stars, exotic stars, wormholes and black holes are astrophysical
objects described by the Einstein's Field equations. For a perfect fluid and
in case of spherical symmetry, these objects obey the
Tolman-Oppenheimer-Volkoff (TOV) equation (in c.g.s. units) \cite{Tolman,OV}
\begin{equation}
\frac{dp_{r}\left(  r\right)  }{dr}=-\left(  \rho\left(  r\right)
+\frac{p_{r}\left(  r\right)  }{c^{2}}\right)  \frac{4\pi Gr^{3}p_{r}\left(
r\right)  /c^{2}+Gm(r)}{r^{2}\left[  1-2Gm(r)/rc^{2}\right]  }+\frac{2}%
{r}\left(  p_{t}\left(  r\right)  -p_{r}\left(  r\right)  \right)  \label{TOV}%
\end{equation}
and
\begin{equation}
\frac{dm}{dr}=4\pi\rho\left(  r\right)  r^{2},
\end{equation}
where $c$ is the velocity of light, $G$ is the gravitational constant,
$\rho\left(  r\right)  $ is the macroscopic energy density measured in proper
coordinates, $p_{r}\left(  r\right)  $ and $p_{t}\left(  r\right)  $ are
respectively the radial pressure and the transverse pressure and $m(r)$ is an
arbitrary function of the radial coordinate, $r$. The function $m(r)$ is the
quasi-local mass, and is denoted as the mass function. It is clear that the
knowledge of $\rho\left(  r\right)  $ allows to understand the astrophysical
structure under examination. If we fix our attention on compact stars,
ordinary General Relativity offers two kind of exact solutions for the
isotropic TOV equation:

a) the constant energy density solution,

b) the Misner-Zapolsky energy density solution\cite{MZ}

or the combination of a) and b), namely the Dev-Gleiser energy density
profile\cite{DG}. Since compact stars are usually macroscopic objects, the
Quantum Gravity contribution is expected to become important when the inner
core of the star is considered, where the highest pressures and densities are
reached. An attempt to include quantum gravitational effects in compact stars,
besides those that are consequences of the standard Fermi degeneracy pressure,
can be found in\cite{Cam}, where Planck scale modifications of the
energy/momentum dispersion relations have been taken into the account, and in
\cite{WYZ}, where the TOV equation and the equation of state of zero
temperature ultra-relativistic Fermi gas based on generalized uncertainty
principle (GUP) have been used to see the quantum gravitational effects on the
cores of compact stars. Gravity's Rainbow offers another opportunity to probe
quantum gravitational effects into the core of a compact star. For simplicity
we will fix our attention only on the isotropic case.

\section{Gravity's Rainbow and the Equation of State}

Basically, Gravity's Rainbow is a distortion of space-time induced by two
arbitrary functions, $g_{1}\left(  E/E_{\mathrm{Pl}}\right)  $ and
$g_{2}\left(  E/E_{\mathrm{Pl}}\right)  $, which have the following property%
\begin{equation}
\lim_{E/E_{\mathrm{Pl}}\rightarrow0}g_{1}\left(  E/E_{\mathrm{Pl}}\right)
=1\qquad\text{and}\qquad\lim_{E/E_{\mathrm{Pl}}\rightarrow0}g_{2}\left(
E/E_{\mathrm{Pl}}\right)  =1.\label{lim}%
\end{equation}
It has been introduced for the first time by Magueijo and Smolin\cite{MagSmo},
who proposed that the energy-momentum tensor and the Einstein's Field
Equations were modified with the introduction of a one parameter family of
equations\footnote{Applications and implications of Gravity's Rainbow in
Astrophysics and cosmology can be found in\cite{GRw}}
\begin{equation}
G_{\mu\nu}\left(  E/E_{\mathrm{Pl}}\right)  =8\pi G\left(  E/E_{\mathrm{Pl}%
}\right)  T_{\mu\nu}\left(  E/E_{\mathrm{Pl}}\right)  +g_{\mu\nu}%
\Lambda\left(  E/E_{\mathrm{Pl}}\right)  ,\label{Gmn}%
\end{equation}
where $G\left(  E/E_{\mathrm{Pl}}\right)  $ is an energy dependent Newton's
constant and $\Lambda\left(  E/E_{\mathrm{Pl}}\right)  $ is an energy
dependent cosmological constant, defined so that $G\left(  0\right)  $ is the
low-energy Newton's constant and $\Lambda\left(  0\right)  $ is the low-energy
cosmological constant. For instance, the \textit{rainbow} version of the
Schwarzschild line element is
\begin{equation}
ds^{2}=-\left(  1-\frac{2MG\left(  0\right)  }{r}\right)  \frac{d\tilde{t}%
^{2}}{g_{1}^{2}\left(  E/E_{\mathrm{Pl}}\right)  }+\frac{d\tilde{r}^{2}%
}{\left(  1-\frac{2MG\left(  0\right)  }{r}\right)  g_{2}^{2}\left(
E/E_{\mathrm{Pl}}\right)  }+\frac{\tilde{r}^{2}}{g_{2}^{2}\left(
E/E_{\mathrm{Pl}}\right)  }d\Omega^{2},\label{line}%
\end{equation}
where $d\Omega^{2}=d\theta^{2}+\sin^{2}\theta d\phi^{2}$ is the line element
of the unit sphere. It is immediate to generalize the metric $\left(
\ref{line}\right)  $ for any spherically symmetric spacetime%
\begin{equation}
ds^{2}=-\frac{e^{2\Phi(r)}}{g_{1}^{2}(E/E_{\mathrm{Pl}})}c^{2}dt^{2}%
+\frac{dr^{2}}{g_{2}^{2}(E/E_{\mathrm{Pl}})\left(  1-\frac{2Gm(r)}{rc^{2}%
}\right)  }+\frac{r^{2}}{g_{2}^{2}(E/E_{\mathrm{Pl}})}d\Omega^{2},\label{dS}%
\end{equation}
where $m(r)$ is the mass of the star inside the radius $r$ and $\Phi(r)$ is
the redshift function. Of course, the line element $\left(  \ref{dS}\right)  $
has consequences on Eq.$\left(  \ref{TOV}\right)  $. To see what are these
consequences, we consider the energy-momentum stress tensor describing a
perfect-fluid of the form%
\begin{equation}
T_{\mu\nu}=\left(  \rho\left(  r\right)  c^{2}+p_{t}\right)  u_{\mu}u_{\nu
}+p_{t}g_{\mu\nu}+\left(  p_{r}-p_{t}\right)  n_{\mu}n_{\nu},
\end{equation}
where $u^{\mu}$ is the four-velocity normalized in such a way that $g_{\mu\nu
}u^{\mu}u^{\nu}=-1$, $n_{\mu}$ is the unit spacelike vector in the radial
direction, i.e. $g_{\mu\nu}n^{\mu}n^{\nu}=1$ with $n^{\mu}=\sqrt{1-2Gm\left(
r\right)  /rc^{2}}\delta_{r}^{\mu}$. $\rho\left(  r\right)  $ is the energy
density, $p_{r}\left(  r\right)  $ is the radial pressure measured in the
direction of $n^{\mu}$, and $p_{t}\left(  r\right)  $ is the transverse
pressure measured in the orthogonal direction to $n^{\mu}$. Because of
Gravity's Rainbow, the normalization of $u^{\mu}$ is modified and becomes%
\begin{equation}
-1=-\frac{e^{2\Phi(r)}}{g_{1}^{2}(E/E_{\mathrm{Pl}})}u^{0}u^{0}\rightarrow
\quad u^{0}=g_{1}(E/E_{\mathrm{Pl}})e^{-\Phi(r)},
\end{equation}
and for $n^{\mu}$, one gets%
\begin{equation}
1=\frac{n^{1}n^{1}}{g_{2}^{2}(E/E_{\mathrm{Pl}})\left(  1-2Gm\left(  r\right)
/rc^{2}\right)  }\rightarrow\quad n^{1}=g_{2}(E/E_{\mathrm{Pl}})\sqrt
{1-2Gm\left(  r\right)  /rc^{2}}.
\end{equation}
Fixing our attention on the isotropic case, the Stress-Energy tensor becomes%
\begin{equation}
\begin{aligned}[l] T_{00} & =\frac{\rho\left( r\right) c^{2}e^{2\Phi(r)}}{g_{1}^{2}(E/E_{\mathrm{Pl}})}c^{2}\\ T_{22} & =\frac{pr^{2}}{g_{2}^{2}(E/E_{\mathrm{Pl}})}\\ \end{aligned}\hspace
{2cm}%
\begin{aligned}[l] T_{11} & =\frac{p(r)}{g_{2}^{2}(E/E_{\mathrm{Pl}})\left[ 1-2Gm\left( r\right) /rc^{2}\right] }\\ T_{33} & =\frac{pr^{2}\sin^{2}\theta}{g_{2}^{2}(E/E_{\mathrm{Pl}})}.\label{stress} \end{aligned}%
\end{equation}
and the component of the Einstein tensor $G_{00}$ reduces to%
\begin{equation}
G_{00}=2G\frac{e^{2\Phi(r)}}{r^{2}}\frac{g_{2}^{2}(E/E_{\mathrm{Pl}})}%
{g_{1}^{2}(E/E_{\mathrm{Pl}})}m^{\prime}(r).\label{G00}%
\end{equation}
With the help of the first component of the Stress-Energy tensor $\left(
\ref{stress}\right)  $ and Eq.$\left(  \ref{G00}\right)  $, we can write the
first Einstein's Field Equation, namely $G_{00}=\kappa T_{00}$ which assumes
the form%
\begin{equation}
m^{\prime}(r)=\frac{\kappa\rho(r)r^{2}}{c^{2}g_{2}^{2}(E/E_{\mathrm{Pl}})},
\end{equation}
while for the second one, namely $G_{11}=\kappa T_{11}$, we get%
\begin{equation}
\Phi^{\prime}(r)=\frac{\kappa r^{3}p_{r}/g_{2}^{2}(E/E_{\mathrm{Pl}}%
)c^{4}+2Gm(r)/c^{2}}{2r^{2}\left[  1-2Gm(r)/rc^{2}\right]  }.\label{eq:Pot-1}%
\end{equation}
It is important to say that the equilibrium equation
\begin{equation}
\frac{dp}{dr}+\left(  \epsilon+p\right)  \Phi^{\prime}(r)=0\label{EqEq}%
\end{equation}
is not affected by Gravity's Rainbow. From Eq.$\left(  \ref{EqEq}\right)  $,
it follows that
\begin{equation}
\frac{dp_{r}}{dr}=-\left(  \rho+\frac{p_{r}}{c^{2}}\right)  \frac{\kappa
r^{3}p_{r}/g_{2}^{2}(E/E_{\mathrm{Pl}})c^{4}+2Gm(r)/c^{2}}{2r^{2}\left[
1-2Gm(r)/rc^{2}\right]  },\label{TOVGRw}%
\end{equation}
and
\begin{equation}
\frac{dm}{dr}=\frac{4\pi\rho(r)r^{2}}{g_{2}^{2}(E/E_{\mathrm{Pl}}%
)},\label{m'(r)}%
\end{equation}
where $\rho$ is the mass density. Eq.$\left(  \ref{TOVGRw}\right)  $
represents the TOV equation modified by Gravity's Rainbow. We will fix our
attention on the constant energy density case and to the variable case of the
Misner-Zapolsky type.

\subsection{Isotropic pressure and the constant energy density case}

The constant energy density case, represents the simplest case to consider.
With this assumption, equation $\left(  \ref{TOVGRw}\right)  $ becomes
\begin{equation}
\frac{dp_{r}}{dr}=-\left(  \rho+\frac{p_{r}(r)}{c^{2}}\right)  \frac{4\pi
Gr^{3}p_{r}(r)/c^{2}g_{2}^{2}(E/E_{\mathrm{Pl}})+Gm(r)}{r^{2}\left[
1-2Gm(r)/rc^{2}\right]  },\label{RTOV2}%
\end{equation}
while Eq.$\left(  \ref{m'(r)}\right)  $ can be easily solved to give
\begin{equation}
m(r)=\frac{4\pi\rho}{3g_{2}^{2}(E/E_{\mathrm{Pl}})}r^{3},\label{eq:mg2}%
\end{equation}
where we have used the boundary condition $m(0)=0$. It is important to observe
that the mass density is constant in $r$, but it is not constant in $E$. It is
also important to observe that Eqs.$\left(  \ref{RTOV2}\right)  $ and $\left(
\ref{eq:mg2}\right)  $ work for the whole star included the external boundary
$R$, where we can assume that the effects of Gravity's Rainbow have vanished.
To this purpose, we analyze the problem into two fundamental
regions\cite{RGGM}:

\begin{itemize}
\item[a)] The boundary $R\ggg\alpha l_{\mathrm{Pl}}$, namely the boundary is
very large compared to the size of the inner core.

\item[b)] The boundary $R\simeq\alpha l_{\mathrm{Pl}}$, that it means that we
are exploring the possibility of the existence of stars of Planckian size. It
is interesting to note that both cases respect the Buchdahl-Bondi bound which
states that\cite{BB}
\begin{equation}
M<\frac{4}{9}\frac{c^{2}}{G}R.\label{BB}%
\end{equation}

\end{itemize}

The case $b)$ can be interpreted as a star forming close to the Planck scale
and stabilized by Gravity's Rainbow. This means that it is the distorted
space-time which supports the existence of a star of Planckian size.

\subsection{Isotropic pressure and the variable energy density case}

The variable energy density case is represented by the Misner-Zapolsky
solution\cite{MZ}. To discuss the modification induced by Gravity's Rainbow,
we consider a density energy profile of the following form
\begin{equation}
\rho=Ar^{\alpha},
\end{equation}
where $A$ is a constant with dimensions of an energy density divided by a
(length)$^{\alpha}$ with $\alpha\in\mathbb{R}$ to be determined. Solving
Eq.$\left(  \ref{m'(r)}\right)  $ leads to
\begin{equation}
m(r)=\int_{0}^{r}\frac{4\pi A}{g_{2}^{2}(E/E_{\mathrm{Pl}})}r^{\prime2+\alpha
}dr^{\prime}=\frac{4\pi A}{g_{2}^{2}(E/E_{\mathrm{Pl}})\left(  3+\alpha
\right)  }r^{3+\alpha}.\label{m(r)}%
\end{equation}
Plugging $\left(  \ref{m(r)}\right)  $ into Eq.$\left(  \ref{TOVGRw}\right)
$, one finds
\begin{gather}
\omega\frac{d\rho\left(  r\right)  }{dr}=-\rho\left(  r\right)  \left(
\frac{c^{2}+\omega}{c^{2}}\right)  \frac{4\pi Gr^{3}\omega\rho(r)+Gm(r)c^{2}%
g_{2}^{2}(E/E_{\mathrm{Pl}})}{r^{2}\left[  1-2Gm(r)/rc^{2}\right]  c^{2}%
g_{2}^{2}(E/E_{\mathrm{Pl}})}\nonumber\\
\Downarrow\\
\alpha=-\left(  \frac{c^{2}+\omega}{\omega c^{2}}\right)  \frac{4\pi
GAr^{2+\alpha}\left(  \left(  3+\alpha\right)  \omega+c^{2}\right)  }{\left[
c^{2}g_{2}^{2}(E/E_{\mathrm{Pl}})\left(  3+\alpha\right)  -8\pi GAr^{2+\alpha
}\right]  },
\end{gather}
where we have used the following Equation of State $p_{r}(r)=\omega\rho\left(
r\right)  $. It is immediate to see that $\forall\alpha\neq-2$, there is a
singularity into the TOV equation and a dependence on $r$ still persists.
Therefore fixing $\alpha=-2$ one gets the relationship
\begin{equation}
1=\frac{3\left(  c^{2}+\omega\right)  ^{2}}{4\omega\left[  7c^{2}g_{2}%
^{2}(E/E_{\mathrm{Pl}})-3\right]  },\label{omega}%
\end{equation}
where we have set $A=3c^{2}/\left(  56\pi G\right)  $. We find an identity
when $\omega=1/3$, $\omega=3$, $c=1$ and $g_{2}(E/E_{\mathrm{Pl}})=1$, namely
we get the ordinary GR solution of the undeformed TOV Equation. In particular
for $\omega=1/3$
\begin{equation}
p_{r}=\omega\rho\left(  r\right)  =\omega\frac{3c^{2}}{56\pi Gr^{2}}%
=\frac{c^{2}}{56\pi Gr^{2}}\label{MZ}%
\end{equation}
and
\begin{equation}
m(r)=\frac{3\pi c^{2}r}{14G},
\end{equation}
we reproduce the Misner-Zapolsky solution. On the other hand, when Gravity's
Rainbow is switched on and $g_{2}(E/E_{\mathrm{Pl}})\neq1$, it is immediate to
see that from Eq.$\left(  \ref{omega}\right)  $ follows that $\omega$ is no
longer a constant but it becomes a function of $E/E_{\mathrm{Pl}}$.

\section{Summary and further comment}

In this work, we have considered the possibility that a compact star is
affected by Gravity's Rainbow. Since the action of Gravity's Rainbow is
prevalently at Planckian length scales, we find that in case of isotropic
pressure and constant energy density, a star of Planckian size if it is
formed, and satisfies the usual Buchdahl-Bondi bound, is also stable. On the
other hand, when the variable energy density case is considered and an
equation of state is introduced, one finds that, from the relation
$p_{r}=\omega\rho\left(  r\right)  $, $\omega$ becomes a function of
$E/E_{\mathrm{Pl}}$, necessarily. It is interesting to note that the constant
energy density and the Misner-Zapolsky energy density are two particular cases
of the Dev-Gleiser potential which is of the form\cite{DG}%
\begin{equation}
\rho\left(  r\right)  =\rho_{0}+\frac{A}{r^{2}},
\end{equation}
where $\rho_{0}$ is the parameter of the constant energy density case and
$A=3c^{2}/\left(  56\pi G\right)  $. Note that in both cases, namely the
constant and variable energy density case, also the mass becomes a function of
$E/E_{\mathrm{Pl}}$. Here we have considered the simple case where
$E/E_{\mathrm{Pl}}$ is not dependent on the radius $r$. Of course, other than
introducing an anisotropy, the case in which $E/E_{\mathrm{Pl}}$ becomes
$E\left(  r\right)  /E_{\mathrm{Pl}}$ will be a subject of a future
investigation as well as the full examination of the Dev-Gleiser potential.

\end{document}